# Zero Knowledge Proof based authentication protocol using graph isomorphism

Lavish Saluja
Department of Electrical and Electronics Engineering
Birla Institute of Technology and Science, Pilani - Pilani
Pilani, India
lavishsaluja.ls@gmail.com

Ashutosh Bhatia
Department of Computer Science and Information Systems
Birla Institute of Technology and Science, Pilani - Pilani
Pilani, India
ashutosh.bhatia@pilani.bits-pilani.ac.in

**ABSTRACT:**

We live in an era of information and it is very important to handle the exchange of information. While sending data to an authorized source, we need to protect it from unauthorized sources, changes, and authentication. ZKP technique can be used in designing secure authentication systems which don't involve any direct exchange of information between the claimant and the verifier thus preventing any possible leak of personal information.

We propose a Zero-Knowledge Proof (ZKP) algorithm based on isomorphic graphs. We suggest most of the computations should be carried out on user's web browser without revealing the password to the server at any point of time. Instead, it will generate random graphs and their permutations based on the login ID and password.

*Keywords*

Zero-Knowledge Proof, Graph Isomorphism, Password,

## I. INTRODUCTION:

Each time an application provides authentication, some data (password and login) are sent from the client to the servers through HTTP (in most cases), though for security reasons they are not stored in plain text form and given to the servers in a readable form. For commercial software, developers use public key cryptographic protocols such as HTTPS. It sets up secure communication between the client and server but the login and password details are still sent.

The motivation for this research is to design, develop and implement a fast authentication model using zero-knowledge proof to provide added security.

PAPER OVERVIEW:

Zero-Knowledge Proof is explained in **Section II**. In **Section III**, we explain the authentication process and in **Section IV**, we describe the protocol and discuss the various private key algorithm in **Section V**. We discuss the possible attacks in **Section VI**. We review the related work in this area in **Section VII**. We conclude the paper in **Section VIII**.

## II. ZERO-KNOWLEDGE PROOF:

Canonical interactive proof systems have two parties: a prover and a verifier. [1]

Zero-Knowledge Proof according to the Wikipedia definition[1] is a method by which one party (the prover) can prove to the other party (the verifier) that they know a value x, without revealing any other information than the fact that they know the value x.

Now we are going to apply the same concept while authenticating users by not revealing their passwords to the server (the verifier) and still be able to prove that the client has the password.

In the ZKP protocol, the verifier can't learn anything from the authentication process. The prover can't cheat on the verifier to be someone else because of repeated challenge response action and the verifier can't pretend to be prover to a third party because the verifier will always have only one value which will always be insufficient to find or calculate the prover's secret.

We discuss different approaches/problems to implement zero-knowledge proof. Each of them will have different complexities, use different data structures and hence have unique properties.

CLASSICAL PROBLEMS

The two classical problems that enable zero-knowledge proof are discrete logarithm problem (DLP)[2] and square root problem (SRP). Both of them depends on generating prime numbers.

Elliptical Curve Cryptography (ECC)[3] is an attractive alternative to the discrete logarithm or square root problem or methods such as RSA. ECC is an approach to public-key cryptography based on the algebraic structure of elliptical curves over finite fields. Also, ECC is an ideal candidate for the constrained devices where the major computational source (memory, speed) are limited. The majority of operations in ECC involves point multiplication. It becomes less accurate because of the errors while multiplications and in order to make them accurate, the points are represented using Jacobi coordinates[4]. In this way, we can eliminate the field inversion in the scalar multiplications. The principle of the optimization done being the point doubling happens faster than the point addition in Jacobi coordinates. It eliminates the need to calculate many multiplication inverses but it requires more scalar multiplications with affine coordinates. We'll skip going into much details over ECC here though since it's out of the scope of this paper but there has been quite an extensive research over it.

---

[1] Wikipedia - ZKP
[2] DLP
[3] ECC
[4] Jacobi_coordinates

## GRAPH ISOMORPHISM

Two graphs $G_1$ and $G_2$ are said to be isomorphic, if a one-to-one permutation or mapping exist between the set of vertices of $G_1$ and the set of vertices of $G_2$ with the property that if two nodes of $G_1$ are adjacent, so are their images in $G_2$. Now it's not an easy task to develop a practical algorithm to find if two given graphs are isomorphic or not but for some simple classes of graphs (such as planar), there exists such efficient algorithms. Whereas, for hard graphs, there exist algorithms with exponential upper bound time. However, the *nauty* package developed by Brendan McKay[5] is known to be the most efficient algorithm. [2] [3] [4] [5] [6]

Suppose there are two graphs $G_1$ and $G_2$ such that the graph $G_2$ is generated by relabeling the vertices of graph $G_1$. The permutation used will be the secret key and the graphs $G_1$ and $G_2$ will be the public key pair. Another random permutation will be used to generate a third graph H which will be sent to the verifier which will in return challenge the prover to provide a permutation which can map H back with either of the graphs $G_1$ or $G_2$. So the procedure for doing the zero-knowledge proof using graph isomorphism goes as:

1. Given $G_1$ and $G_2$ such that $G_2 = \pi(G_1)$
2. Prover randomly chooses a number out of {1,2} and call it "a".
3. Prover chooses a random permutation $\varepsilon$ over $G_a$ and generates another graph H where $H = \varepsilon(G_a)$.
4. Prover sends the adjacency matrix of H to the verifier.
5. Verifier in return sends a number "b" randomly from the set {1,2}.
6. If a = b, the prover sends $\chi = \varepsilon^{-1}$ to the verifier.
7. If a =1 and b = 2, the prover sends $\chi = \varepsilon^{-1} \circ \pi$ to the verifier.
8. If a = 2 and b = 1, the prover sends $\chi = \varepsilon^{-1} \circ \pi^{-1}$ to the verifier.
9. Verifier checks if $\chi(H) = G_b$ and grants access to the prover.

There is a probability of ½ of prover to be lucky and get access even without having the right permutation (secret key) and thus this several rounds of this method are needed for the verifier to get completely convinced with the prover's identity. After ten rounds, the confidence level of the verifier is approximately 99.99%.

This Graph Isomorphism based Zero-Knowledge protocol definitely has the three must properties:
1. Soundness, *since the verifier follows the protocol.*
2. Completeness: *since both parties follow the protocol.*
3. Zero Knowledge: *since the verifier doesn't learn anything other than the validity of the proof provided by the verifier.*

### III. AUTHENTICATION PROCESS:

The Classical way [6] of authenticating users is asking for their login and password details and sending these details directly to the server and the server responds to the request based on the validity of login details. But the credentials are still sent and are given to the servers in readable form even if they are hashed or salted. Both SSL[7] (Secure sockets layer) and TLS (Transport layer security) can provide the client, server and annual entity authentication.

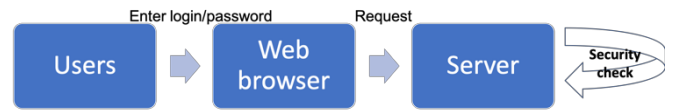

Figure 1: Authentication on the web

The most common usage for digital certificates is entity authentication when attempting to connect through a secure website (SSL). For high security systems, client side authentication[8] is a must. Installing a certificate to user's web browser happens rarely and is prone to risks. In the end, to take the full use of HTTPS and stay safe, the users should buy a certificate issued by a well-known certificate authority and then the certificate should be installed on user's browser.

### IV. THE ZKP APPROACH FOR AUTHENTICATION:

Now, with the ZKP the verifier (server) can't be convinced with the login and password form, therefore, the ZKP approach is different from the classical approaches as mentioned in Section III and thus requires more extensive computations, bandwidth, flexible communication means.

We tried replacing the extensive computations (like calculating co-prime numbers or calculating multiplication inverses, etc.) by choosing a NP problem and using asynchronous web technologies like Ajax.

A diagram presented in Figure 2 shows our approach in detail. The user types her name and password but the password would never leave her browser unlike in the case the classical way of authentication. The browser uses the login and password details to generate a secret key (permutation π) and the public key pair ($G_1$ and $G_2$).

---

[5] http://users.cecs.anu.edu.au/~bdm/

[6] Web authentication
[7] SSL, TSL, HTTPS
[8] Client side authentication

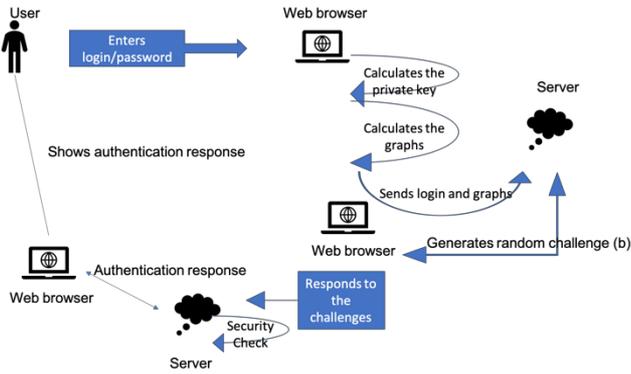

Figure 2: ZKP Implementation on the web

| Classical | Zero Knowledge Proof |
| --- | --- |
| -- | Generate private key |
| -- | Generate challenging graphs |
| Send login, password | Send login, graphs |
| -- | Responds to challenges. |
| Receive and show an authentication response. | Receive and show authentication response |

Table 1: Compares the work of browser in classical and ZKP based authentication systems in web.

### V. PRIVATE KEY ALGORITHM:

The private key here is the permutation $\pi$ that the browser generates on the basis of login and password details. We are using one way functions over standard secure hashing algorithms (SHA)[9] to generate corresponding permutations to the password. A hash function [10] is a mathematical operation which maps data of any size (password in our case) to a bit string of fixed size. As stated by OWASP, hash functions have the following properties:

1. It's easy and practical to compute the hash but it's impossible to regenerate the only input with the hash value.
2. It's very difficult to design an input to match a desired output.

### VI. POSSIBLE ATTACKS:

The implementation can be done in a few ways, either as a browser's extension, as a script or as a third party website. If the server can't be trusted scripting gives no benefit and developing an extension would be beneficial but will require more efforts from users. We have assumed there is no phishing like sign-in-seal. [2]

Attacks on the approach include our dependence on different algorithms and functions used like:

1. **SHA-1:**

---
[9] Secure hash algorithm
[10] Hashing passwords - one way to security

We are using the secure hash algorithm (SHA-1) which is supposed to need $2^{80}$ operations for a phishing attack which was later brought down to $2^{63}$ operations as presented in August 2005 attack.

2. **Secure communication**:
We are assuming that the communication between user and her browser is completely secure and thus there are no vulnerabilities, cross site scripting, any malicious software, etc.

3. **Dictionary Attacks:**
Florencio[8] conducted a research about password habits. Such kind of attacks happen because of poor password choice by the users which are easily guessable or are reused at many websites.[11] Now, If the user's and server's communication is encrypted the only way to attack is to run a dictionary which can be slowed down and detected as proved already using captchas.
A new protocol was developed to counter online dictionary attacks which suggested using account locking, delayed response, use of CAPTCHA.[12][13]

4. **NP Problem:**
Efficient algorithms for some classic graphs were proposed for determining if two graphs are isomorphic. However, in our approach the main focus is to find the permutation $\pi$ and not to prove if two graphs are isomorphic. The generated graphs are completely random and none of the proposed algorithms can be used to find out the secret.

### VII. RELATED WORK:

I came across NORWAHL, a research output by Ryan Cheu, MIT. They developed and deployed a zero knowledge proof based authentication system and used discrete logarithms as their main problem to be used in challenge response action.

Yahoo![9] proposed a simple challenge response solution a long ago, which was very similar to HTTP Digest. Because of major possibilities of servers to impersonate user even after hashing password, this was labelled as quite insecure.

Encrypted Key exchange [7] (EKE) was introduced in 1992, which takes both asymmetric and symmetric cryptographic results. It uses one modulo function.

Secure Remote password (SRP) was developed in 1997 whose security was directly proportional to the security of applied one way hashing function. It was quite computationally expensive because eit uses two modulo functions. It was quite vulnerable to dictionary attacks.

### VIII. CONCLUSION AND FUTURE WORK:

We discussed and presented a novice way of authenticating users in the web by using graph isomorphism based Zero-Knowledge Proof protocols that is sound, complete and give zero knowledge to the server regarding user's password.

We discussed how the ZKP based authentication protocols give non trivial benefits over the classical approaches.

We also discussed about the computations that should be done on user's browser in order to improve the accuracy and the complexity of the process.

We plan to work more on the implementation and use JavaScript to implement the client side extension and use asynchronous web technologies along with graph isomorphism which would be implemented using the popular *nauty* package. The next stepts would be enabling easy, reusable system.

Further we could also implement one of the protocols that secures against the targeted attack described above in Section VI. The weakness aside, the GIZKP still stands out as an effective authentication system.